\documentclass[fp]{jpsj3}
\usepackage{txfonts}
\usepackage{color}

\title{
Valence Change Driven by Constituent Element Substitution in the Mixed-Valence Quasicrystal and Approximant Au-Al-Yb   
}

\author{
Shuya Matsukawa$^1$, Katsumasa Tanaka$^2$, Mika Nakayama$^1$, Kazuhiko Deguchi$^1$, Keiichiro Imura$^1$, Hiroyuki Takakura$^2$, Shiro Kashimoto$^2$, Tsutomu Ishimasa$^2$, and Noriaki K. Sato$^1$\thanks{E-mail: kensho@cc.nagoya-u.ac.jp}
}

\inst{
$^1$Department of Physics, Graduate School of Science, Nagoya University, Nagoya 464-8602, Japan \\
$^2$Division of Applied Physics, Graduate School of Engineering, Hokkaido University, Sapporo 060-8628, Japan 
}

\abst{
Quantum criticality has been considered to be specific to crystalline materials such as heavy fermions.
Very recently, however, the Tsai-type quasicrystal Au$_{51}$Al$_{34}$Yb$_{15}$ has been reported to show unusual quantum critical behavior.
To obtain a deeper understanding of this new material, we have searched for other Tsai-type cluster materials.
Here, we report that the metal alloys Au$_{44}$Ga$_{41}$Yb$_{15}$ 
and Ag$_{47}$Ga$_{38}$Yb$_{15}$ are
members of the 1/1 approximant to the Tsai-type quasicrystal
and that both possess no localized magnetic moment.
We suggest that the Au-Al-Yb system is located near the border of the divalent and trivalent states of the Yb ion;
we also discuss a possible origin of the disappearance of magnetism, associated with the valence change, by the substitution of the constituent elements.
}

\begin{document}
\maketitle

\section{Introduction}

\begin{figure}[t]
\begin{center}
\includegraphics[scale=0.55]{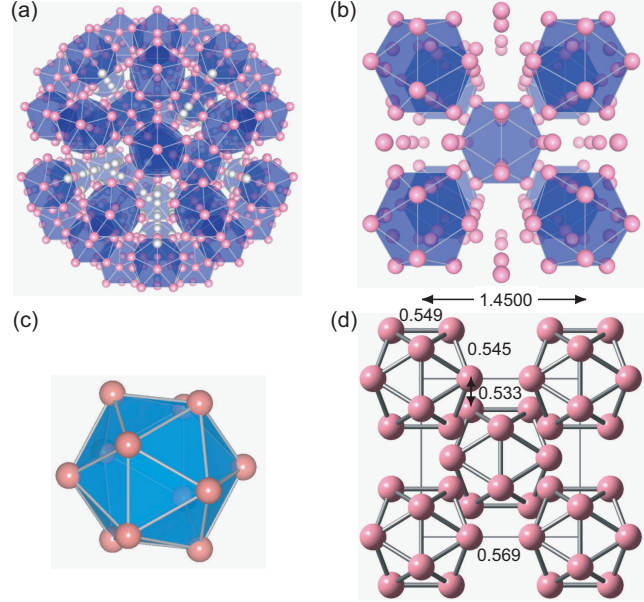}
\end{center}
\caption{(Color)
Geometric structure of 
the Tsai-type quasicrystal (QC) and approximant crystal (AC).
(a) Aperiodic array of icosahedrons (denoted by blue polyhedron) in QC.
Rose-pink and gray spheres denote Yb 
ions located on the vertices of an icosahedron [Fig. \ref{fig_0}(c)] and in an acute rhombohedron [see Fig. \ref{fig_I2}(b)], respectively.
These two Yb sites are inequivalent and called the A- and B-sites throughout this paper.
(b) Body-centered cubic array of the icosahedrons in AC. 
(c) Icosahedron consisting of 12 Yb ions. 
(d) Network of Yb ions in Au-Al-Yb 1/1 AC. The number denotes the length in the unit of nm. 
}
\label{fig_0}
\end{figure}

Quasicrystals (QCs) are metallic alloys that have long-range, aperiodic structures with diffraction symmetries forbidden to conventional crystals [see Fig. \ref{fig_0}(a)].~\cite{Schechtman} 
Because of their unique structure,
one may expect the existence of a critical state that is neither extended nor localized.
This novel electronic state has long been pursued by experimentalists
but has not been established yet.
Very recently,
a quantum critical phenomenon characterized by the divergence of the magnetic susceptibility ($\chi \propto T^{-0.51}$) and electronic specific heat coefficient ($C/T \propto -\ln T$, where $C$ is the specific heat) as $T \rightarrow 0$,
was observed in the Au$_{51}$Al$_{34}$Yb$_{15}$ QC,~\cite{Deguchi} in the course of research on a new series of Tsai-type QCs.~\cite{Ishimasa}
This observation has twofold implications.
First, 
the observed quantum criticality can be a sign of the QC critical state 
because the approximant crystal [AC, see Fig.~\ref{fig_0}(b)] shows no such divergence ($\chi^{-1} \propto T^{0.51} + {\rm constant}$);~\cite{Deguchi}
note that AC is a phase whose unit cell has atomic decorations [i.e., icosahedral clusters of atoms shown in Fig.~\ref{fig_0}(c)] 
that look like the QC.
Second, 
a study of this new type of QC may shed more light on the basic notions of quantum criticality
because the critical indices of the novel QC are very similar to those observed in some Yb-based heavy fermions.~\cite{Y1,Y2,Y3,Y4,Watanabe} 

The geometric structures of the Au-Al-Yb QC and AC can be understood using the structure models of the Cd-$R$ ACs ($R$=rare-earth element).~\cite{4,5}
The icosahedrons are arranged quasiperiodically 
with a fivefold diffraction symmetry in the QC
[Fig.~\ref{fig_0}(a)], 
while in the 1/1 AC, 
they are arranged periodically to form a body-centered cubic (bcc) structure 
(space group: $Im\bar{3}$) [Fig.~\ref{fig_0}(b)].~\cite{6}
This bcc structure of the icosahedrons
may be more clearly seen in Fig. \ref{fig_0}(d);
each icosahedron has edge lengths of 0.545 and 0.549 nm, 
and the interatomic distance between the neighboring icosahedrons is 0.533 and 0.569 nm. 
These distances are so large that there is no direct overlap between the 4$f$ electron wave functions derived from the Yb ions.
%
\begin{figure*}[b]
\begin{center}
\includegraphics[scale=0.8]{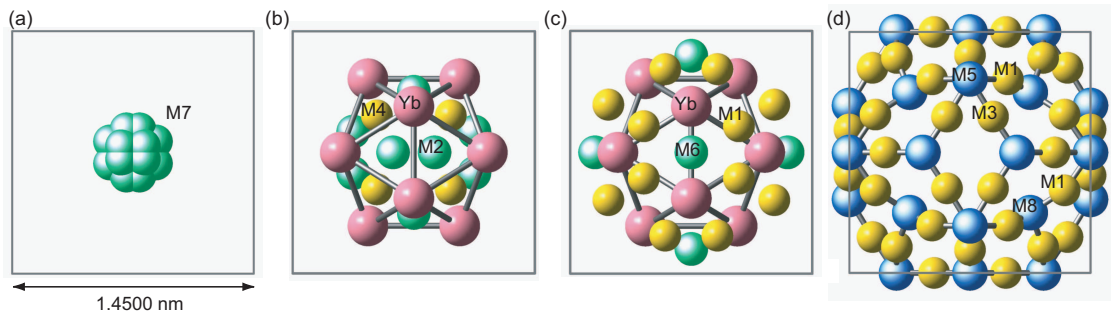}
\end{center}
\caption{(Color)
Structure model of Tsai-type cluster in Au-Al-Yb 1/1 approximant. Rose-pink: Yb, blue: Al, yellow: Au, green: Au/Al. 
The square frame of each panel indicates the unit cell size of the lattice parameter $a=1.4500$ nm.
(a) Local structure at the center of the cluster.
The tetrahedron is oriented in a disordered fashion, and the average of the various orientations leads to the complicated polyhedron structure.
(b) First and second shells of the cluster.
(c) Second and third shells of the cluster.
(d) Triacontahedron decorated by Au and Al atoms. 
}
\label{fig_I1}
\end{figure*}
%
%
In contrast to information on such geometric structures,
that on the electronic structure of the QC and AC is as yet limited;
although it was reported that the Yb ion of the QC and AC is in a mixed-valence state of Yb$^{2+}$ and Yb$^{3+}$,~\cite{Deguchi,Watanuki} 
it remains unknown if the mixed valence nature and the unusual criticality are unique to the Au-Al-Yb system.
In this research, we study Au-Ga-Yb and Ag-Ga-Yb systems to obtain a deeper understanding of the electronic state of the Au-Al-Yb systems. We find that both Au-Ga-Yb and Ag-Ga-Yb alloys have the Tsai-type cluster
and show a nonmagnetic state with the ionic state of Yb$^{2+}$; 
we also discuss a possible origin of the disappearance of magnetism by the substitution of the constituent elements in the Au-Al-Yb system.
\begin{table}[t]
\caption{
Summary of information on site M$_{i}$ ($i=$1-7) 
contained in the Yb coordination polyhedron shown in Fig. \ref{fig_I2}(a).~\cite{Ishimasa} 
The second column denotes the color of the spheres shown in Figs. 1-3. y: yellow, g: green, b: blue. 
The third column indicates the number of atoms belonging to the M$_{i}$ site. 
The fourth column indicates the probability of each site being occupied by Al.
Note that the partially occupied site M7 is occupied by 1/6 atoms on average; 8.9\% by Al and 7.8\% by Au. 
The last column indicates the distance between the Yb ion and the ligand ion at site M$_{i}$.
For example, there are six ions at the M$_{1}$ site. Two of them are 0.314 nm, two are 0.313 nm, and two are at 0.310 nm from the Yb ion. 
These are abbreviated as 0.310 - 0.314 nm.
}
\label{table2}
\begin{center}
\begin{tabular}{ccccc}
\hline
\multicolumn{1}{c}{Site} & \multicolumn{1}{c}{Color} & \multicolumn{1}{c}{Number} & \multicolumn{1}{c}{Occupation by Al (\%)} & \multicolumn{1}{c}{Distance (nm)}\\
\hline
M1 & y & 6 & 27.2 & 0.310 - 0.314 \\
M2 & g & 3 & 62.0 &  0.311 - 0.328 \\
M3 & y & 3 & 11.8 &  0.319 - 0.322 \\
M4 & y & 2 & 4.4 & 0.316 \\
M5 & b & 1 & 96.7 & 0.331\\
M6 & g & 1 & 58.9 & 0.307 \\
M7 & g & 1/6 $\times$ 2 & 8.9 & 0.389 \\
\hline
\end{tabular}
\end{center}
\end{table}

The remaining part of this section is dedicated to a more detailed description of the geometric structure of the QC and AC, which will help us to explore the relationship between the geometric and electronic structures.
Figure \ref{fig_I1} shows a structure model of the Au-Al-Yb 1/1 AC determined by the Rietveld method.
The square frame indicates the unit cell size of a lattice parameter $a$ = 1.4500 nm. 
The Tsai-type cluster consists of a concentric arrangement of triple shells (see ref.~2 for the shell arrangement);
at the center of the first shell, there is a polyhedron labeled M7 [Fig.~\ref{fig_I1}(a)]. 
Here, the green sphere denotes the site alternatively occupied by Au and Al (see Table \ref{table2}). 
This polyhedron consists of four atoms in total (abbreviated as 4 Au/Al atoms in this paper) and is represented by a complicated polyhedron reflecting an average of variously oriented tetrahedrons.~\cite{6}. 
Figure \ref{fig_I1}(b) shows the first and second shells, in which one may find a pentagon beneath the rose-pink sphere (denoted Yb) consisting of three green spheres (M2) and two yellow spheres (M4).
Here, the rose-pink sphere denotes the site occupied by Yb,
and the yellow one indicates the site preferably occupied by Au. 
Figure \ref{fig_I1}(c) shows the second and third shells, in which one may find another pentagon [consisting of four yellow spheres (M1) and one green sphere (M6)] around Yb.
In Fig.~\ref{fig_I1}(d), 
we find five yellow spheres (M1 and M3) 
to form a pentagon around the blue sphere (M5);
here, the blue sphere indicates the site almost occupied by Al.

\begin{figure}[b]
\begin{center}
\includegraphics[scale=0.67]{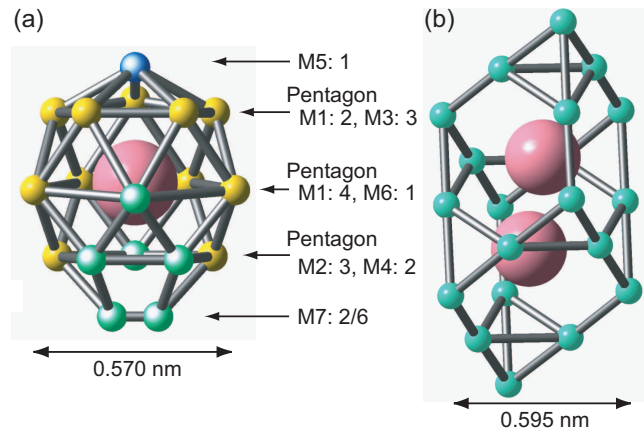}
\end{center}
\caption{(Color)
Geometric arrangement of Yb ions in AC and QC.  
(a) Coordination polyhedron of Yb. (b) Acute rhombohedron in Cd-Yb QC, in which green spheres indicate Cd.
This type of Yb site is missing in 1/1 ACs. 
}
\label{fig_I2}
\end{figure}

The three pentagons together with the blue (M5) and green (M7) spheres are arranged as shown in Fig. \ref{fig_I2}(a). 
This figure may best visualize the local environment of the Yb ion.
In this structure model of AC, the Yb ion site is uniquely determined and referred to as the A-site hereafter. 
By analogy with Cd-$R$ QCs,~\cite{5} on the other hand, 
there are two Yb sites in the Au-Al-Yb QC: one is the A-site and the other is a specific site (referred to as the B-site hereafter) embedded in the so-called acute rhombohedron (AR) [Fig.~\ref{fig_I2}(b)]. 
Each AR includes two B-sites with a distance of about 0.34 nm, which is too large to give rise to a direct overlap between the 4$f$ wave functions. 
Note that the population ratio of the A- and B-sites is approximately 7:3.

According to a nuclear magnetic resonance measurement for the Au-Al-Yb QC and AC,  
nuclear magnetization recoveries after saturation pulses were fit using a theoretical function with a single component of $T_1$ ($T_1$ is a nuclear spin-lattice relaxation time) for both the QC and AC.~\cite{Deguchi} This suggests that the electronic state is rather homogeneous even in the QC.
Therefore, we assume throughout the present study that the Yb sites are homogeneous from the viewpoint of the electronic structure.

\section{Experimental Procedure}

The purities of the starting materials used in the present study were as follows (in wt\% );
Au: 99.99, Ag: 99.999, Al: 99.999, Ga: 99.999, and Yb: 99.9. 
These starting materials were arc-melted on a water-cooled copper hearth under an argon atmosphere to prepare QCs.
ACs were prepared by two methods: the arc-melting of the starting materials and the subsequent annealing of the obtained alloy-ingot in an evacuated quartz ampoule under the conditions summarized in Table \ref{table0},
and the melting of the starting materials in an evacuated quartz ampoule as in annealing.
The Au-Al-Yb AC sample whose data are presented in Figs. \ref{fig:suscep} and \ref{fig:sus1} was prepared by the second method.
Note that the alloy composition is a starting, nominal composition and that the principal magnetic property
(i.e., the absence/presence of the localized moment)
of the samples studied here
is independent of the sample preparation method and annealing conditions.  
\begin{table}[b]
\caption{
Summary of the structure analysis results.
The composition given in the first column is the nominal composition.
$a_{\rm 6D}$ indicates a six-dimensional lattice parameter of the obtained QC.  
}
\label{table0}
\begin{center}
\begin{tabular}{lcc}
\hline
\multicolumn{1}{c}{Alloys} & \multicolumn{1}{c}{Heat treatments} & \multicolumn{1}{c}{Structures}\\
\hline
Au$_{49}$Al$_{34}$Yb$_{17}$ & as-cast & P-type QC, $a_{\rm 6D}$ = 0.7448 (2) nm \\
Au$_{49}$Al$_{36}$Yb$_{15}$ & 700 $^\circ$C, 24 h &  BCC 1/1-AC, $a$ = 1.4500 (2) nm \\
Au$_{44}$Ga$_{41}$Yb$_{15}$ & 650 $^\circ$C, 27 h & BCC 1/1-AC, $a$ = 1.4527 (1) nm \\
Ag$_{47}$Ga$_{38}$Yb$_{15}$ & 503 $^\circ$C, 92 h & BCC 1/1-AC, $a$ = 1.4687 (1) nm \\
\hline
\end{tabular}
\end{center}
\end{table}

Structure characterization was carried out by powder X-ray diffraction analysis as well as by selected-area electron diffraction methods. Powder diffraction patterns were measured using Cu $K\alpha$-radiation. The lattice parameters presented in Table \ref{table0} were determined by Cohen's and the Rietveld method for the QCs and ACs, respectively. 
Selected-area electron diffraction experiments were carried out using a JEOL JEM-200CS microscope operating at 200 kV. 

The magnetization was measured using a commercial magnetometer in terms of a superconducting quantum interference device (SQUID) in the temperature range between 1.8 and 300 K at magnetic fields up to 7 T.

\section{Results and Discussion}

\subsection{Finding of new Tsai-type AC: Au-Ga-Yb}

\begin{figure}[b]
\begin{center}
\includegraphics[scale=0.8]{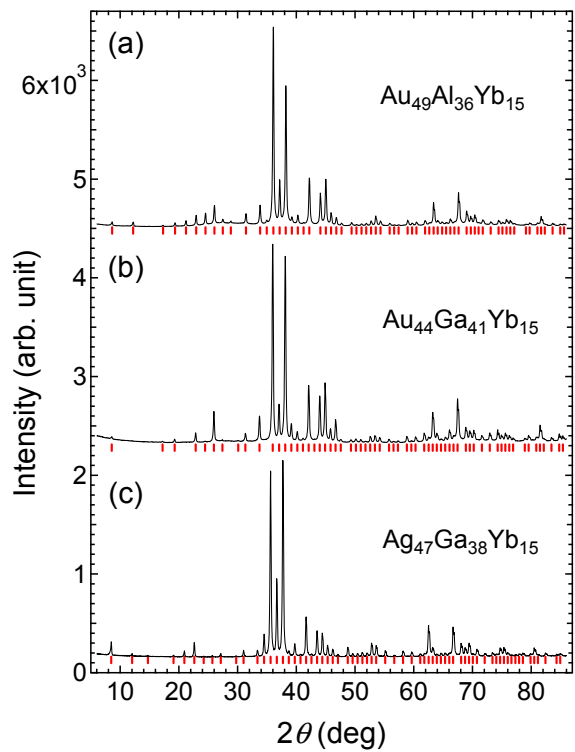}
\end{center}
\caption{(Color online)
Powder X-ray diffraction patterns of ACs formed in annealed alloys. Positions of Bragg reflections are indicated by bars at the lower part of each pattern. (a) Au$_{49}$Al$_{36}$Yb$_{15}$.
(b) Au$_{44}$Ga$_{41}$Yb$_{15}$. (c) Ag$_{47}$Ga$_{38}$Yb$_{15}$.
}
\label{fig_I3}
\end{figure}

Figure \ref{fig_I3} shows the powder X-ray diffraction patterns of the annealed alloys of (a) Au$_{49}$Al$_{36}$Yb$_{15}$, (b)
Au$_{44}$Ga$_{41}$Yb$_{15}$, and (c) Ag$_{47}$Ga$_{38}$Yb$_{15}$.
The spectrum of 
the Au-Al-Yb alloy is given as a reference of the Tsai-type 1/1 AC,
and
the Ag-Ga-Yb alloy is known to form the Tsai-type cluster similar to that of the Au-Al-Yb AC,~\cite{7} 
but the structure of the Au-Ga-Yb alloy has not been reported thus far.
As seen in the figure,
all reflections of the three alloys were indexed by bcc structures as indicated by short vertical bars, with the lattice parameters listed in the third column of Table \ref{table0}. The present Rietveld analysis result suggests that they are of the same structure type,
meaning that the Au-Ga-Yb alloy is a new member of the Tsai-type 1/1 AC.

\subsection{Origin of the crystal field effect in Au-Al-Yb}

\begin{figure}[b]
\begin{center}
\includegraphics[scale=0.83]{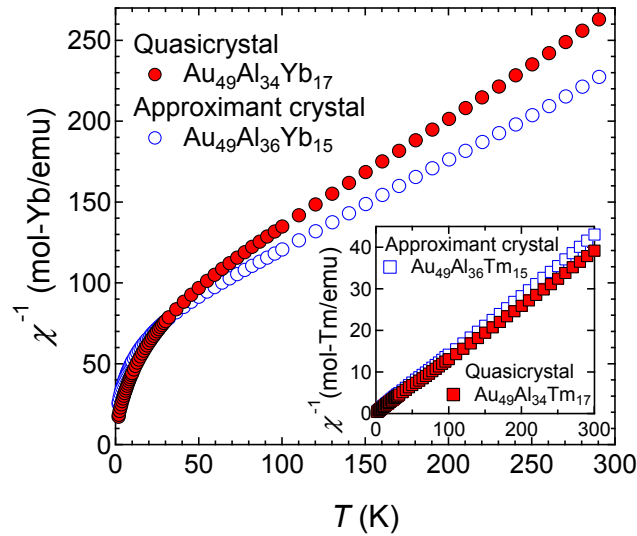}
\end{center}
\caption{(Color online)
Temperature dependences of the inverse magnetic susceptibilities of the Au-Al-Yb QC and AC measured at $H=$ 5 and 1 kOe, respectively.
The inset shows the temperature dependence of the inverse magnetic susceptibility of the Au-Al-Tm systems measured at $H=$ 0.5 kOe.  
Note the very similar behaviors between the QC (Au$_{49}$Al$_{34}$Tm$_{17}$) and AC (Au$_{49}$Al$_{36}$Tm$_{15}$) with slopes (7.8 and 7.4 $\mu_{\rm B}$/Tm, respectively) corresponding to that of a free Tm$^{3+}$ ion.
Detailed results will be published elsewhere.
}
\label{fig:suscep}
\end{figure}

Figure \ref{fig:suscep} shows the temperature dependences of the inverse magnetic susceptibilities of the Au-Al-Yb QC and AC.
They show very similar features, i.e., a straight line feature at high temperatures and a convex curvature below about 50 K.
This indicates that the magnetism in the temperature range presented here is not related to the absence/presence of the periodicity, as shown in Figs.~\ref{fig_0}(a) and \ref{fig_0}(b), but 
rather to the local structure, as shown in Figs.~\ref{fig_I2}(a) and \ref{fig_I2}(b).

The linear slope of the $\chi(T)^{-1}$ curves above $\sim$150 K yields an effective magnetic moment of $p_{\rm eff}=3.44 $ and 3.79 $\mu_{\rm B}$/Yb for the QC and AC, respectively.
These values are smaller than the free ion value of Yb$^{3+}$, 4.54$\mu_{\rm B}$/Yb, confirming the mixed valence nature.
Note that
there is a sample dependence of $p_{\rm eff}$; 
the previously reported values are 3.91 and 3.96 $\mu_{\rm B}$/Yb for the QC and AC, respectively.~\cite{Deguchi} 
At present, it is unclear how this sample dependence is related to the crystal structure parameter such as the occupation probability in Table I.

The extrapolation of the high-temperature linear portion into low temperatures gives a large Weiss temperature $\theta_{\rm P}$, 
defined as $\chi(T) \propto p_{\rm eff}^2/(T + \theta_{\rm P})$,
on the order of 100 K.
We ascribe this to the crystal field effect as is usually done for rare-earth crystalline materials;
a large antiferromagnetic interaction between localized spins is very unlikely as suggested from the absence of antiferromagnetic order down to the base temperature ($\sim $2 K).
What is the origin of the crystal field effect? 
There are two possible origins: the hybridization effect between the Yb and ligand ions, and the Coulomb potential due to the ligand ions.
As the present system is a mixed-valence system, the former origin seems favorable.
To confirm this, we measured the magnetic susceptibility of the Au-Al-Tm systems with the same geometric structure as the Yb systems; the results are plotted in the inset of Fig. \ref{fig:suscep} in the form of $\chi(T)^{-1}$ vs $T$.
The straight line feature over a very wide temperature range with $p_{\rm eff}$ close to that of a free Tm$^{3+}$ ion,
which is in marked contrast with the convex curvature of the Yb systems,
indicates that the 4$f$ electrons of the Tm systems are well localized in the real space.
Note that, in the case of the localized moment system, the crystal field effect should arise from the Coulomb potential.
The dissimilarity between the Tm and Yb systems supports the above possibility that the crystal field effect in the Au-Al-Yb AC is due to the hybridization effect.

\subsection{Effect of substitution of Ga for Al on the magnetism}

\begin{figure}[b]
\begin{center}
\includegraphics[scale=0.84]{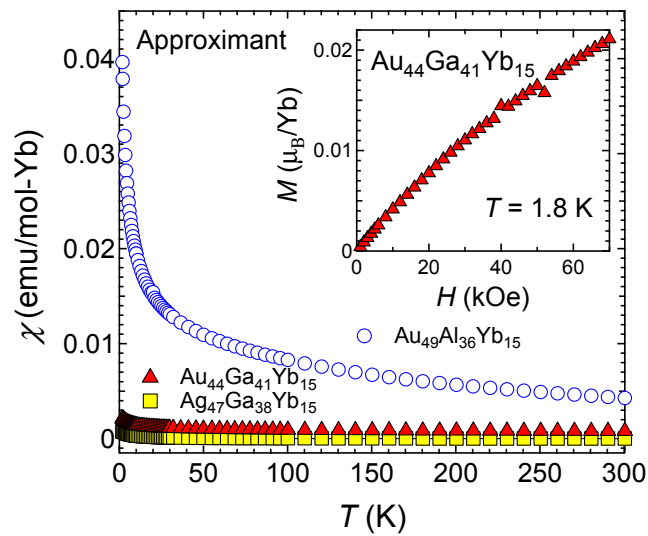}
\end{center}
\caption{(Color online)
(a) Temperature dependences of the magnetic susceptibilities of the Au-Al-Yb, Au-Ga-Yb, and Ag-Ga-Yb ACs.
The magnetization was measured at $H =$ 1, 0.1, and 5 kOe for the Au-Al-Yb, Au-Ga-Yb, and Ag-Ga-Yb ACs, respectively. Note that the magnetism disappears with the substitution of Ga for Al. 
The inset shows the magnetization curve for the Au-Ga-Yb AC at $T=1.8$ K.
}
\label{fig:sus1}
\end{figure}

Figure \ref{fig:sus1} shows the temperature dependences of the magnetic susceptibilities of the Au-Ga-Yb and Ag-Ga-Yb ACs. 
For comparison,
we replotted the susceptibility of the Au-Al-Yb AC shown in Fig. \ref{fig:suscep}.
The Ga-substituted systems for Al exhibit a temperature-independent behavior, indicating that the Yb ion is in the nonmagnetic Yb$^{2+}$ state.
They also show a rise in $\chi(T)$ in the lowest-temperature region measured,
but this seems extrinsic because the $\chi(T)$ and the finite-field magnetization $M(H)$ at a low temperature are very small; $M(H)$ is only 0.02 $\mu_{\rm B}$/Yb at $H =$ 70 kOe (see inset).
As a result,
the mixed-valence nature observed in the Au-Al-Yb AC is not observed in the Au-Ga-Yb and Ag-Ga-Yb ACs.
This suggests that the Au-Al-Yb system is located near the border of the Yb$^{2+}$ and Yb$^{3+}$ ionic states.

Let us discuss the origin of the disappearance of the magnetism associated with the valence change.
As seen in Table II, the lattice of the Au-Ga-Yb and Ag-Ga-Yb AC expands by about 0.2 and 1.3\%, respectively, compared with that of the Au-Al-Yb AC. 
As pressure favors an $f$ configuration with a smaller volume, i.e., Yb$^{3+}$ in the present case,~\cite{Goltsev}
the valence change in Au-Al-Yb $\rightarrow$ Ag-Ga-Yb can be 
naively
interpreted to be a result of an internal, negative pressure.
However, this interpretation may not apply in 
the case of Au-Al-Yb $\rightarrow$ Au-Ga-Yb, 
because 
the lattice parameter change is very small.
Here, we should remember that the pressure effect 
(i.e., the effect of the change of the lattice parameter)
is not the only mechanism responsible for the change in the valence and hence in magnetism.
In general, the appearance/disappearance of the localized moment in a metal may be determined by the relative magnitude of the Coulomb repulsion and the mixing energy $\Delta = \pi |V|^2 \rho(E_{\rm F})$, 
or equivalently, the ratio $|\varepsilon_{f} - E_{\rm F}|/\Delta$;
here, $V$ is a mixing matrix element between the wave functions of 4$f$ and conduction electrons,
$\rho(E_{\rm F})$ is the density of states at the Fermi energy $E_{\rm F}$, 
and $\varepsilon_{f}$ is the energy level of the 4$f$ electron.
Note 
that the substitution of constituent elements, as employed here, can cause a change in the magnitude of $V$ and hence in $\Delta$, and
that the smaller the ratio the closer the Yb valence to 2+.
In the case of Au-Al-Yb $\rightarrow$ Au-Ga-Yb, we conjecture that
the substitution  
would give rise to 
a decrease in the ratio
via
an increase in $V$
assuming that the spatial distribution of the outermost (presumably $p$) electron wave function would be wider for Ga than for Al. 
In the case of Au-Al-Yb $\rightarrow$ Ag-Ga-Yb, on the other hand, the negative pressure will lead to a decrease in both $V$ and $|\varepsilon_{f} - E_{\rm F}|$, and $|\varepsilon_{f} - E_{\rm F}|$ should become small faster than $\Delta$, as suggested by the disappearance of the localized moment.
Band structure calculation will be helpful for confirming this possibility.

Next, we discuss the relationship between the geometric and electronic structures. 
As shown in the atomic coordination structure of Fig. \ref{fig_I2}(a),
the distances between the Yb site and the ligand ion M$_i$ site are almost the same, and the probability of the M$_i$ site being occupied by Al depends on the site (see Table \ref{table2}).
It seems reasonable to assume that the occupation probabilities are not very different between the Au-Al-Yb and Au-Ga-Yb systems, and that
the M5 site with the highest occupation-probability of Al is very susceptible to the effect of the replacement of Al.
Then,
we speculate that the mixing between the Yb ion and the Al/Ga ion at the M5 site
would play an important role in the magnetism of the present systems;
it is beyond our scope to determine whether this model explains the magnetism observed here.

Finally, we discuss the magnetism of the QC.
We have not succeeded in preparing the Au-Ga-Yb QC; hence, we have no information on its magnetic properties.
However,
the Au-Ga-Yb QC will also exhibit a nonmagnetic behavior similarly to the Au-Ga-Yb AC because high-temperature magnetism is not affected by the absence/presence of translational symmetry, as discussed above.
As a result,
$p$-$f$ mixing is likely to be a key factor controlling the magnetism of the Au-Al-Yb QC.

\section{Conclusions}

The Au-Al-Yb quasicrystal and crystalline 1/1 approximant exhibit a mixed valence nature. More interesting is the unusual quantum criticality that the quasicrystal shows at very low temperatures. 
To gain a deeper understanding of the electronic structure of these systems, we synthesized two materials; Au$_{44}$Ga$_{41}$Yb$_{15}$ and Ag$_{47}$Ga$_{38}$Yb$_{15}$.
We observed that annealed alloys of both these materials belong to a Tsai-type 1/1 crystalline approximant 
and exhibit lattice expansion by about 0.2 and 1.3\% compared with the Au-Al-Yb approximant crystal, respectively.
We further observed that they exhibit a nonmagnetic behavior 
corresponding to that of Yb$^{2+}$.
From these results, 
we suggested that the Au-Al-Yb approximant crystal is located near the border of the Yb$^{2+}$ and Yb$^{3+}$ ionic states.
We then
discussed a possible origin of the changes in the valence and magnetism of Au-Al-Yb $\rightarrow$ Au-Ga-Yb and Ag-Ga-Yb in terms of the ratio $|\varepsilon_{f} - E_{\rm F}|/\Delta$.

For the Au-Ga-Yb quasicrystal,
we have no experimental information on its magnetism. 
However, it would most likely be nonmagnetic because the high-temperature magnetism is not affected by the presence/absence of translational symmetry.
This allows us to 
conjecture
that the effect of hybridization between the Yb and Al ions 
plays a role in
the mixed valence nature in the Au-Al-Yb quasicrystal as in the Au-Al-Yb approximant crystal.

In conclusion, the Au-Al-Yb quasicrystal and approximant crystal are, at present, the only mixed-valence systems at ambient pressure.
We hope that the present study stimulates further search for new alloy systems with a mixed valence.

\begin{acknowledgment}

The authors thank Y. Tanaka and S. Yamamoto for experimental support, and
S. Watanabe, K. Miyake, and K. Hattori for valuable discussions. 
This work was partially supported by grants-in-aid for Scientific Research from JSPS, KAKENHI (Nos. 24654102, 20224015, 20102006, and 25610094).

\end{acknowledgment}

\end{document}